\documentclass[aps,twocolumn,showpacs,superscriptaddress]{revtex4-1}  % for review and submission
\usepackage{graphicx}  % needed for figures
\usepackage{dcolumn}   % needed for some tables
\usepackage{bm}        % for math
\usepackage{amssymb}   % for math
\usepackage{amsmath}   % for math
\usepackage{color}
\usepackage[latin1]{inputenc}

\newcommand{\ket}[1]{\left\vert #1 \right\rangle}
\newcommand{\tr}[1]{\mathrm{tr}\left\{ #1 \right\}}

\newcommand{\dd}{\mathrm{d}}

\newcommand{\fuc}{\tfrac{1}{4}}

\newcommand{\al}{\alpha}
\newcommand{\be}{\beta}
\newcommand{\ga}{\gamma}

\newcommand{\upa}{\uparrow}
\newcommand{\doa}{\downarrow}

\newcommand{\pan}[1]{\left\langle #1 \right\rangle}
\newcommand{\pac}[1]{ \left\{ #1 \right\} }
\newcommand{\pap}[1]{\left( #1 \right)}
\newcommand{\pas}[1]{\left[#1 \right]}

\newcommand{\pav}[1]{\left\vert #1 \right\vert}

\newcommand{\Oc}{\mathcal{O}}
\newcommand{\Sc}{\mathcal{S}}

\newcommand{\So}{\hat{S}}

\newcommand{\Oo}{\hat{O}}
\newcommand{\so}{\hat{s}}
\newcommand{\sbo}{\hat{\mathbf{s}}}

\newcommand{\sio}{\hat{\sigma}}
\newcommand{\Ho}{\hat{H}}

\newcommand{\rv}{\mathbf{r}}
\newcommand{\ev}{\mathbf{e}}

\newcommand{\roo}{\hat{\rho}}
\newcommand{\Ooc}{\hat{\mathcal{O}}}

\begin{document}

\title{Exploring many body localization and thermalization using semiclassical methods}
%RN I think this work is more about the method than the particular signatures, so I changed the title accordingly. (I also dropped `discrete phase space' in the interests of brevity)

\author{O.~L. Acevedo}
\email{osac0358@jilau1.colorado.edu}
\affiliation{JILA, NIST, and Department of Physics, University of Colorado, Boulder, CO 80309, U.S.A.}
\affiliation{Center for Theory of Quantum Matter, University of Colorado, Boulder, CO 80309, U.S.A.}
\author{A. Safavi-Naini}
\affiliation{JILA, NIST, and Department of Physics, University of Colorado, Boulder, CO 80309, U.S.A.}
\affiliation{Center for Theory of Quantum Matter, University of Colorado, Boulder, CO 80309, U.S.A.}
\author{J. Schachenmayer}
\affiliation{JILA, NIST, and Department of Physics, University of Colorado, Boulder, CO 80309, U.S.A.}
\affiliation{Center for Theory of Quantum Matter, University of Colorado, Boulder, CO 80309, U.S.A.}
\affiliation{IPCMS (UMR 7504) and ISIS (UMR 7006), University of Strasbourg and CNRS, 67000 Strasbourg, France}
\author{M. L. Wall}
\thanks{Present address: The Johns Hopkins University Applied Physics Laboratory, Laurel, MD 20723, U.S.A.}
\affiliation{JILA, NIST, and Department of Physics, University of Colorado, Boulder, CO 80309, U.S.A.}
\affiliation{Center for Theory of Quantum Matter, University of Colorado, Boulder, CO 80309, U.S.A.}
\author{R. Nandkishore}
\affiliation{Department of Physics, University of Colorado, Boulder, CO 80309, U.S.A.}
\affiliation{Center for Theory of Quantum Matter, University of Colorado, Boulder, CO 80309, U.S.A.}
\author{A. M. Rey}
\affiliation{JILA, NIST, and Department of Physics, University of Colorado, Boulder, CO 80309, U.S.A.}
\affiliation{Center for Theory of Quantum Matter, University of Colorado, Boulder, CO 80309, U.S.A.}

\begin{abstract}
The Discrete Truncated Wigner Approximation (DTWA) is a semi-classical phase space method useful for the exploration of Many-body quantum dynamics. In this work we investigate Many-Body Localization (MBL) and thermalization using DTWA and compare its performance to exact numerical solutions. By taking as a benchmark case a 1D random field Heisenberg spin chain with short range interactions, and by comparing to numerically exact techniques, we show that DTWA is able to reproduce dynamical signatures that characterize both the thermal and the MBL phases. It exhibits the best quantitative agreement at short times deep in each of the phases and larger mismatches close to the phase transition. The DTWA captures the logarithmic growth of entanglement in the MBL phase, even though a pure classical mean-field analysis would lead to no dynamics at all. Our results suggest the DTWA can become a useful method to investigate MBL and thermalization in experimentally relevant settings intractable with exact numerical techniques, such as systems with long range interactions and/or systems in higher dimensions.
\end{abstract}


\maketitle

\section{Introduction}

Since the seminal discovery of Anderson localization \cite{Anderson1958} it has been known that the transport of an isolated quantum particle can be suppressed by the presence of disorder. While the conditions of this disorder driven localization have become well established for non-interacting particles
\cite{Lee1985}, our understanding of localization of interacting systems is still in rapid development (see \cite{Nandkishore2015} for a recent review). This latter phenomenon, now known as many-body localization (MBL) \cite{Mirlin, BAA, PalHuse}, has gained significant interest motivated by its fundamental connection to failure of thermalization in closed systems \cite{PalHuse, Prosen, Canovi2011,Cohen2016}, emergent integrability \cite{Serbyn, HNO, Scardicchio, Imbrie, lstarbits, GBN}, new types of quantum order with possible applications for improved quantum memories \cite{LPQO, Pekkeretal2014, VoskAltman2014}, due to its novel response properties \cite{nonlocal, mblconductivity} and entanglement structures \cite{BardarsonPollmanMoore, GNR}, and due to the emergence of experimental platforms for investigation of such physics \cite{kondov, Schreiber2015,Monroe2016, Choi2016}.\\

In spite of intensive effort, our current ability to numerically simulate MBL systems faces severe limitations. In particular, currently numerical techniques consist mostly of exact diagonalization, the strong disorder renormalization group \cite{Vosk, VHA, Potter2015}, and variants of the density matrix renormalization group (DMRG) \cite{PekkerClark, KhemaniDMRG}, and are largely limited to one dimensional systems with short range interactions. Given the experimental and theoretical interest in higher dimensional systems and/or systems with long range interactions \footnote{We note that it has been argued \cite{deroeck} that MBL may {\it only} exist in one dimension with short range interactions. Even if this argument is correct, however, the obstruction to MBL comes from `rare regions' and only manifests itself on asymptotically long length scales and timescales, and so the quantum dynamics of disordered systems in higher dimensions and/or with longer range interactions remains experimentally relevant and theoretically interesting}, there is a pressing need for more widely applicable numerical tools for exploring these more general situations \cite{Altman2015}.\\

A recently developed semi-classical phase space method known as Discrete Truncated Wigner Approximation (DTWA) may help to overcome these limitations, as it has been been shown to correctly capture many aspects of the non-equilibrium behavior of many-body quantum spin lattices \cite{Schachenmayer2015,Schachenmayer2015a,Pucci2016,Pucci2017,orioli}. As DTWA is based on solving the system time evolution by classical mean field equations, it has the advantage that its computational complexity increases only %linearly
polynomially with system size. Also, DTWA is not slowed down by the growth of quantum entanglement, which can hinder, for instance, DMRG methods. Given these advantages, we seek to investigate the utility of DTWA for addressing questions regarding many body localization and thermalization. In this work, we benchmark DTWA by studying the 1D nearest neighbor Heisenberg model in random uniaxial field, which is one of the most thoroughly studied systems exhibiting MBL (see e.g.~\cite{Alet2015}). We focus on dynamical signatures of localization that are both suitable to be computed by DTWA, and have also proven experimentally useful in characterizing MBL \cite{Schreiber2015, Monroe2016}. We note that, insofar as DTWA is a `mean field' method, one dimensional systems with short range interactions are expected to be the least favorable setting for this method. Nevertheless, upon benchmarking DTWA to exact diagonalization, we find that DTWA works not just in the short time limit (when it is formally exact), but even at moderately long times. At intermediate times, DTWA is able to reproduce the main dynamical features of both the MBL phase {\it and} the thermal phase. The agreement persists up to longer times as one goes deeper into either phase. We attribute this to both extremes being in some sense classical - the deep MBL regime because there is very little entanglement, and the strongly delocalized regime because there is so much entanglement that reduced density matrices look thermal \cite{Deutsch, Srednicki, Rigol}. We also note that DTWA performs better for some diagnostics than others, and discuss under what circumstances we expect the method to faithfully capture the quantum dynamics.\\

This paper is organized as follows. In Sec. \ref{secmodel} we introduce the model studied and the dynamical quantities that characterize its MBL behavior. In Sec. \ref{secDTWA} we explain the semiclassical phase-space DTWA method. In Sec. \ref{Results}, we present and discuss our results, concluding with some final remarks in Sec. \ref{secconc}.

\section{Spin model and MBL signatures}
\label{secmodel}

In the following we study the dynamical properties of the 1D Heisenberg model with nearest-neighbor interactions and a random uniaxial field described by the Hamiltonian
\begin{equation}\label{eqHam}
\Ho=J\sum_{i=1}^{N-1}\sbo_{i}\cdot\sbo_{i+1}+\sum_{i=1}^{N}h_{i}\so_{i}^{z}\:,
\end{equation}
where $\so_{i}^{\al}=\tfrac{1}{2}\sio_{i}^{\al}$ are spin 1/2 operators ($\sio_{i}^{\al}$ are Pauli matrices), $J$ is the interaction strength, and the values of the random potential $h_i$ (measured in units of $J$) are uniformly distributed $h_i\in\pas{-h,h}$ with $h$ the disorder strength. Previous studies have shown that this model features an MBL transition at $h_c \approx 3.5$ in the thermodynamic limit and at infinite temperature \cite{Huse2010}. At finite temperature the eigenstates exhibit a localization edge between $h \approx 2$ and $h \approx 3.5$ \cite{Alet2015,Abanin2015}.\\

Here we study the dynamics when the system is initialized in the N\'{e}el state, $\ket{\Psi(0)}=\ket{\upa\doa\upa\doa\ldots}$. For this setting, the critical point of $h_c \approx 3.5$ should be taken only as rough reference value. For instance, in the thermodynamic limit the N\'{e}el state is expected to localize below the infinite temperature critical point $h_c$ due to its average energy eigenstates crossing the localization edge at lower disorder strengths. In addition, finite sizes effects may affect the disorder strengths at which signatures of the MBL transition are observed \cite{Potter2015,Pollmann2014}.\\

We probe the dynamics by initializing the system in a Néel state and by then computing dynamical observables which have been previously used to characterize the phase \cite{Huse2010,Moore2012,Pollmann2014,Vasseur2015,Schreiber2015,Monroe2016}. These include the imbalance $I(t)$, Quantum Fisher Information (QFI) denoted by $\mathcal F(t)$, R\'enyi entropy $S_2(t)$ over a two-spin subsystem, and entanglement (Von Neumann) entropy $S_1(t)$ of one half of the system. The imbalance is a measure of the Néel state's staggered order across the spin chain. It is defined as $I\pap{t}=\sum_{i} (-1)^{i+1}\pan{\sio_{i}^z \pap {t}}$. The QFI associated to an operator $\Ooc$ for a spin chain in a pure state is given by $\mathcal{F}_{\Ooc}\equiv4\pap{\pan{\Ooc^{2}}-\pan{\Ooc}^{2}}=4\Delta \Ooc$. Here $\Ooc=\sum_{i=1}^{N}\Oo_{i}$, and each $\Oo_{i}$ is a spin $1/2$ operator acting on spin $i$. The QFI is a witness of multi-partite entanglement if $\mathcal{F}_{\Ooc}/N>1$ \cite{Smerzi2012,TTh2014,Apellaniz2017}. Due to our particular choice of the initial state, it is natural to choose the imbalance as the operator $\Ooc$, so that QFI is proportional to its variance $\mathcal{F}(t)=4\Delta I\pap{t}$. Finally, both R\'enyi entropy $S_2$ and the von-Neumann entanglement entropy $S_1$ can be thought as belonging to a continuous family of entropy measures defined as,
\begin{equation}\label{eqent}
S_{\ga}\pap{\roo}=\frac{1}{1-\ga}\log_2\pas{\tr{\roo^{\ga}}}\:,
\end{equation}
where $\ga \ge 1$. The Rényi entropy corresponds to $S_{\ga=2}$, while the von-Neumann entropy $S_1=\lim_{\ga \to 1}S_{\ga}$. Given that the total spin $\So_z$ is a conserved quantity, and the Néel state is an eigenstate of $\So_z$, the Rényi entropy over a two-spin subsystem with spins labeled as $i$ and $j$ can be particularly expressed as,
\begin{equation}\label{eqrenyi2}
S_{2}\pap{\roo_{i,j}}=-\log_2\pac{\fuc\pas{1+g_{z,0}^{2}+g_{0,z}^{2}+g_{z,z}^{2}+8\pav{g_{+,-}}^{2}}}\:,
\end{equation}
where $g_{\al,\be}=\pan{\sio_{i}^{\al}\sio_{j}^{\be}}$ are two particle moments of Pauli matrices ($\sio^{0}$ is the identity matrix).

\section{Semi-classical phase-space method}
\label{secDTWA}
The DTWA approximately describes the dynamics of quantum observables as statistical averages over classical phase-space trajectories \cite{Schachenmayer2015,Schachenmayer2015a}. It starts from a phase-space formulation of quantum mechanics. The method was inspired by the well known Truncated Wigner Approximation derived for systems with continuous degrees of freedom, described e.g.~by the canonical $\pap{q,p}$ coordinates along each dimension \cite{Polko2010}. In a continuous phase-space, the state of the system (the density matrix $\roo$) and any observable $\Oo$ are represented by real functions of phase-space coordinates, $W(q,p)=\pap{\roo}_W$ and $\Oc (q,p)=\pap{\Oo}_W$ respectively, and any expectation value is computed as,
\begin{equation}
\pan{\Oo}=\tr{\roo \Oo}=\int\!\!\!\!\int\!\!\dd p \dd q \, W(q,p) \Oc (q,p,t)\:.
\label{eqTWAcont}
\end{equation}
The function $W(q,p)$ is known as the Wigner quasi-probability distribution, and the function $\Oc (q,p)$ is known as the Weyl symbol of operator $\Oo$. Either the Wigner function, the Weyl symbol, or both can vary in time depending on the dynamical representation. The previous formula is in the Heisenberg picture, where the operators are the ones that change in time.\\

A similar formulation can be deduced for systems with discrete degrees of freedom, such as the spin chain of $N$ qubits considered in Eq.~\eqref{eqHam} \cite{Wootters1987}. In the discrete case, the phase-space consists of $4^N$ points instead of the continuous $\pap{q,p}$ plane, and a formula analogous to Eq.~\eqref{eqTWAcont} reads,
\begin{equation}\label{eqTWAdisc}
\pan{\Oo\pap t}=\sum_{\mu \in 4^N}w_{\mu}\pap 0\Oc_{\mu}\pap t\:.
\end{equation}
Up to this moment, no approximation has been made: if both $w_{\mu}\pap 0$ and $\Oc_{\mu}\pap t$ are known for every phase-space point, then the previous formula gives the exact quantum solution. The total number of phase-points $\mu$ can become very large, so that a first practical approximation should be done. Any time the Wigner function can be seen as a (positive) probability distribution on phase-space, Eq.~\eqref{eqTWAdisc} can be computed via a Monte-Carlo sampling over a number of phase-space points large enough to obtain an accurate estimate. The initial state in consideration fulfills this positive probability condition.

To compute the Weyl symbols $\Oc_{\mu}\pap t$ efficiently, a more important approximation is required. There is a hierarchy of equations that determine the time evolution of these symbols \cite{Pucci2016}, which can be truncated up to the number of spins in the operators that are considered independent. To first order, i.e. if only single particle operators are considered independent, the equations for $\Oc_{\mu}\pap t$ can be approximated to obey the classical mean field equations,
\begin{equation}\label{eqMFE}
\frac{d\rv_{i,\mu}}{dt}=2J\rv_{i,\mu}\times\pas{\pap{\rv_{i+1,\mu}+\rv_{i-1,\mu}}+h_{i}\ev_{z}}\:,
\end{equation}
where $\rv_{i,\mu}\equiv\pac{\Sc^{\al}_{i}}=\pac{\pap{\sio_{i}^{\al}}_W}$ is the Bloch vector of the $i$-th spin. The index $\mu$ accounts for the specific phase-space point, which in turn defines the initial value of $\rv_{i,\mu}$. Equation~\eqref{eqMFE} is equivalent to assume that all higher order Weyl symbols like $\Sc^{\al,\be}_{i,j}=\pap{\sio_{i}^{\al}\sio_{j}^{\be}}_W$ are approximated as products of single particle Weyl symbols, i.e., $\Sc^{\al,\be}_{i,j} \approx \Sc^{\al}_{i}\Sc^{\be}_{j}$ for $i \neq j$. This approximation should be exact at least at short times for initially uncorrelated spin states as the Néel state in consideration. Thus, the DTWA can be described as a semi-classical method where the time evolution is approximated by classical trajectories whose initial conditions are sampled according to the exact initial Wigner function.\\

It should be noticed that there is a crucial difference between simple mean field solutions and DTWA. In a simple mean field approach, the vectors $\rv_{i,\mu}$ in Eq.~\eqref{eqMFE} represent expectation values and \emph{not} Weyl symbols, and so the set of initial conditions used in Eq.~\eqref{eqMFE} is very different. For the case considered here wherein we start from a N\'{e}el state, a simple mean field approach would predict no dynamics at all. Also, the mean field approach predicts zero correlations between single site observables at any time. Instead, DTWA does develop such correlations in time, further extending its applicability to capture the exact quantum evolution.

\begin{figure}[!t]
  \includegraphics[width=0.49\textwidth]{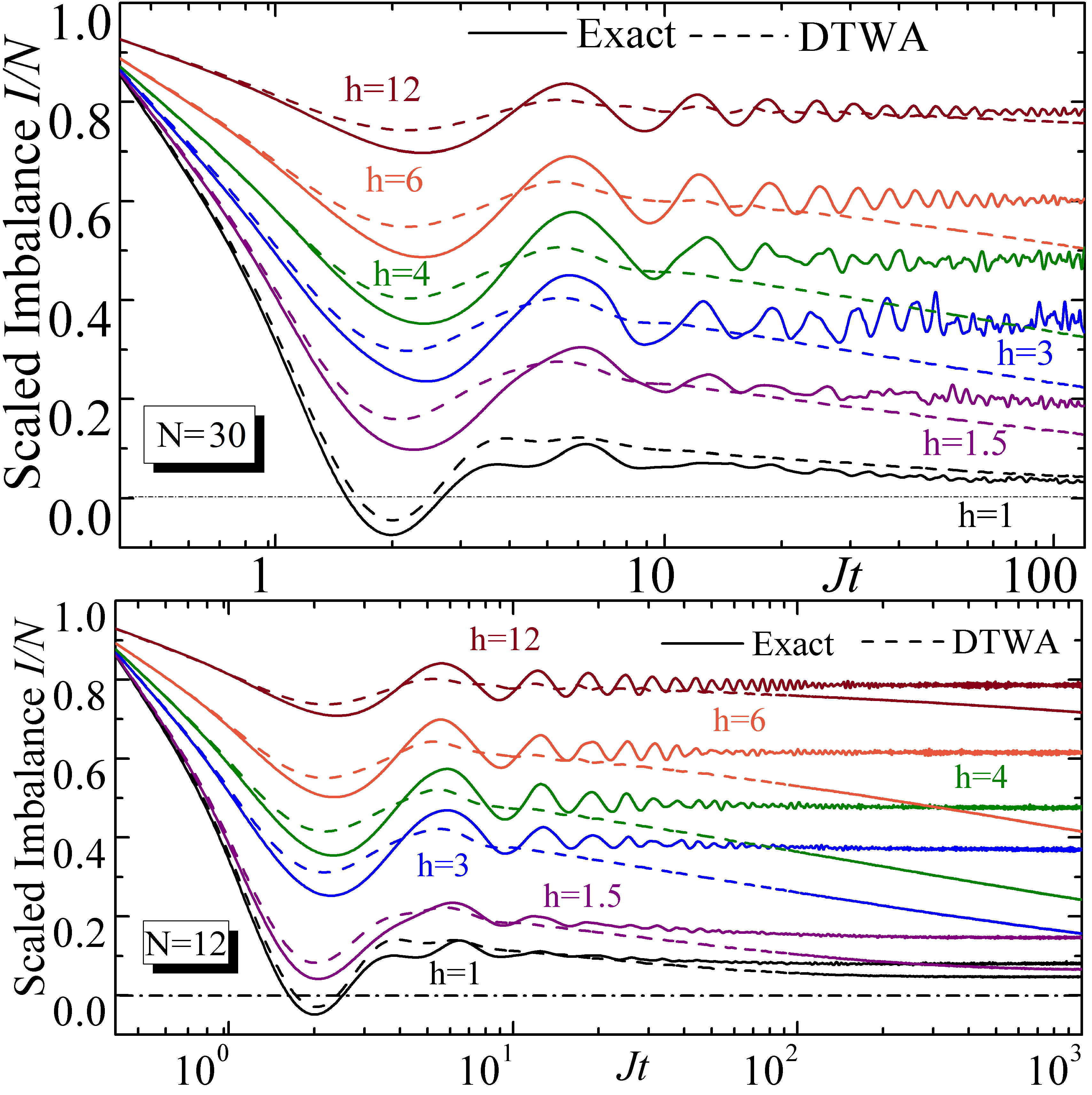}
  \caption{Disorder average of scaled imbalance $I/N$ (initial state order parameter) for system size $N=30$. In the exact results, MBL is manifested when the initial order is partially preserved in the long term as an asymptotic non-zero value. Instead, in the ergodic phase it decays to zero. The cross-over is around $h\approx3$. The semi-classical DTWA results approximately overlap the exact ones at intermediate times, with better agreement for both extremes of disorder strengths. Exact solutions were obtained either by exact diagonalization (bottom), or time-dependent Density Matrix Renormalization Group (t-DMRG) method (top). The bond dimension for the t-DMRG solution was adaptively chosen so that the truncation error never exceeded a tolerance of $10^{-9}$.} \label{figimb}
  \end{figure}

\section{Results}
\label{Results}

In the literature, a commonly used dynamical witness of the MBL phase is the growth of the half chain entanglement entropy $S_{1}\pap{\roo_{N/2}(t)}$ \cite{Moore2012,Altman2015}. However, this quantity is hard to measure experimentally. It is also computationally difficult to estimate within the DTWA. Instead, there are other indicators of localization that are more accessible both to experiments and to the DTWA. One of them is the loss of any initial local structure of the system in the ergodic phase. In the localized phase, on the contrary, there is an extensive set of Local Integrals of Motion (LIOM) that are able to partially preserve the initial order. For instance, in the very large disorder limit, the Pauli operators $\sio_{i}^z$ become a well defined set of LIOM, and any eigenstate of these operators will remain frozen. Therefore, the initial imbalance $I(t)$ remains constant for very large disorder strengths. On the other hand, in the thermal phase, any initial imbalance is expected to decay to zero during the dynamics. At weaker disorder strengths, but still in the localized phase, the LIOM become less local, so that the initial imbalance is no longer a conserved quantity, and cannot stay at its initial value; however, it is still partially conserved and should saturate to a non-zero value in the long term \cite{Huse2010,Vasseur2015}. Figure \ref{figimb} shows this tendency in the exact solutions. We observe a change of behavior around $h\approx3$, which is in close agreement with the critical value expected in the thermodynamic limit \cite{Alet2015,Abanin2015}. The DTWA results are in agreement with the exact dynamics of imbalance at short times, namely $Jt \lesssim 1$. Furthermore, in the intermediate regime $Jt \lesssim 100$ we find good qualitative agreement away from the critical region. We note that the DTWA imbalance decays for all values of disorder strength at sufficiently long times, even deep in the localized phase. This is due to the non-linear nature of the mean field equations used in the DTWA which always leads to long time thermalization, similar to Ref. \cite{Huseclassical}. Nevertheless the imbalance decay is largely suppressed in the localized regime, and the overall slow-down can be used as probe of localization. In fact, the discrepancy with the exact quantum dynamics is delayed further in time as disorder strength increases. In the thermal phase, this decay is real and dictated by the quantum evolution and the DTWA captures well the imbalance dynamics. The agreement in the ergodic phase can be understood by the fact that even though the DTWA neglects higher order correlations between spins, single particle observables like imbalance are not sensitive to those correlations. The role of those higher order correlations when tracing out all but a single spin reduces to a high temperature bath which can also be mimicked by the statistical mixture of classical trajectories. Another difference between the DTWA results and the exact ones are transitory oscillations at intermediate times which are not present in the DTWA solution. These oscillations are fine features generated either by boundary effects or sporadic few-body clusters, and they are not fundamental characteristics of the two phases.\\

\begin{figure}[!t]
  \includegraphics[width=0.49\textwidth]{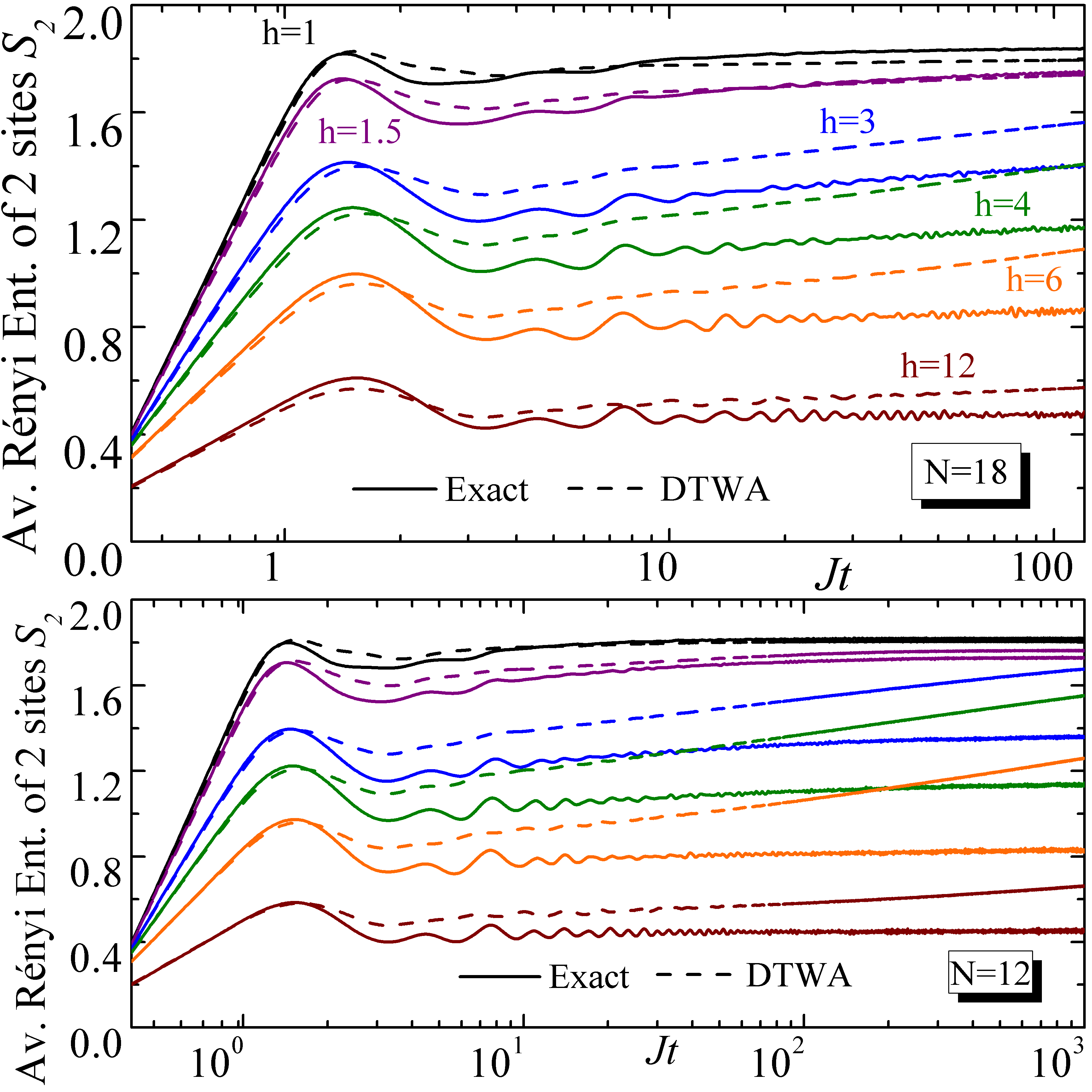}
  \caption{Average Rényi entropy $S_2$ among all possible two site subsystems. System size $N=18$ (top) and $N=12$ (bottom). The MBL regime is characterized by logarithmic growth of $S_2$ even at long times, while the ergodic regime saturates faster. The cross-over is around $h\approx3.5$. The ability of DTWA to reproduce tendencies in the transition regime is similar as in Fig. \ref{figimb}, which is remarkable as this time the observable depends on two sites, and it is related to the entanglement between the subsystem and its surroundings. All exact solutions were obtained by numerical exact diagonalization.} \label{figrent}
  \end{figure}

A similar tendency of agreement between DTWA and exact results is seen for two-site Rényi entropy, as can be seen in Fig. \ref{figrent}. This quantity is a measure of the entanglement between the two-sites-subsystem and the rest of the system. Equation~\eqref{eqrenyi2} shows that $S_2$ is determined by a handful of one and two sites operators. In the long term, both the MBL ($h \ge 3$) and thermal regimes exhibit a monotonic growth of $S_2$, but this growth is slow and logarithmic in the localized regime, while in the thermal regime $S_2$ quickly saturates to a final value. This behavior has strong similarities with known results of entanglement entropy evaluated by tracing over half of the system (see for example \cite{Moore2012}). The main difference is that the upper bound $\max \pap{S_2}=2$ is much lower for a two-spin subsystem. In fact, the saturation values reached by the two-site $S_2$ in the thermal phase are not that different than the ones reached by the same observable in the localized phase. In general terms, the DTWA is able to reproduce the main features exhibited by the different regimes, though it overestimates the logarithmic growth in the MBL phase. As with imbalance, the disparity between the DTWA and the exact results gets smaller as the disorder increases. From Eq.~\eqref{eqrenyi2} one expects that most of the arguments that explain the ability of DTWA to reproduce the imbalance evolution still hold for $S_2$. However, given that the subsystem is composed of two sites, there can be an extra dependence on spin-spin correlations that are not well approximated by DTWA. This explains the discrepancy with the exact results at intermediate times in the localized phase. After thermalization, the spin-spin correlations lose their significance and the agreement with DTWA is recovered.\\

\begin{figure}[!t]
  \includegraphics[width=0.49\textwidth]{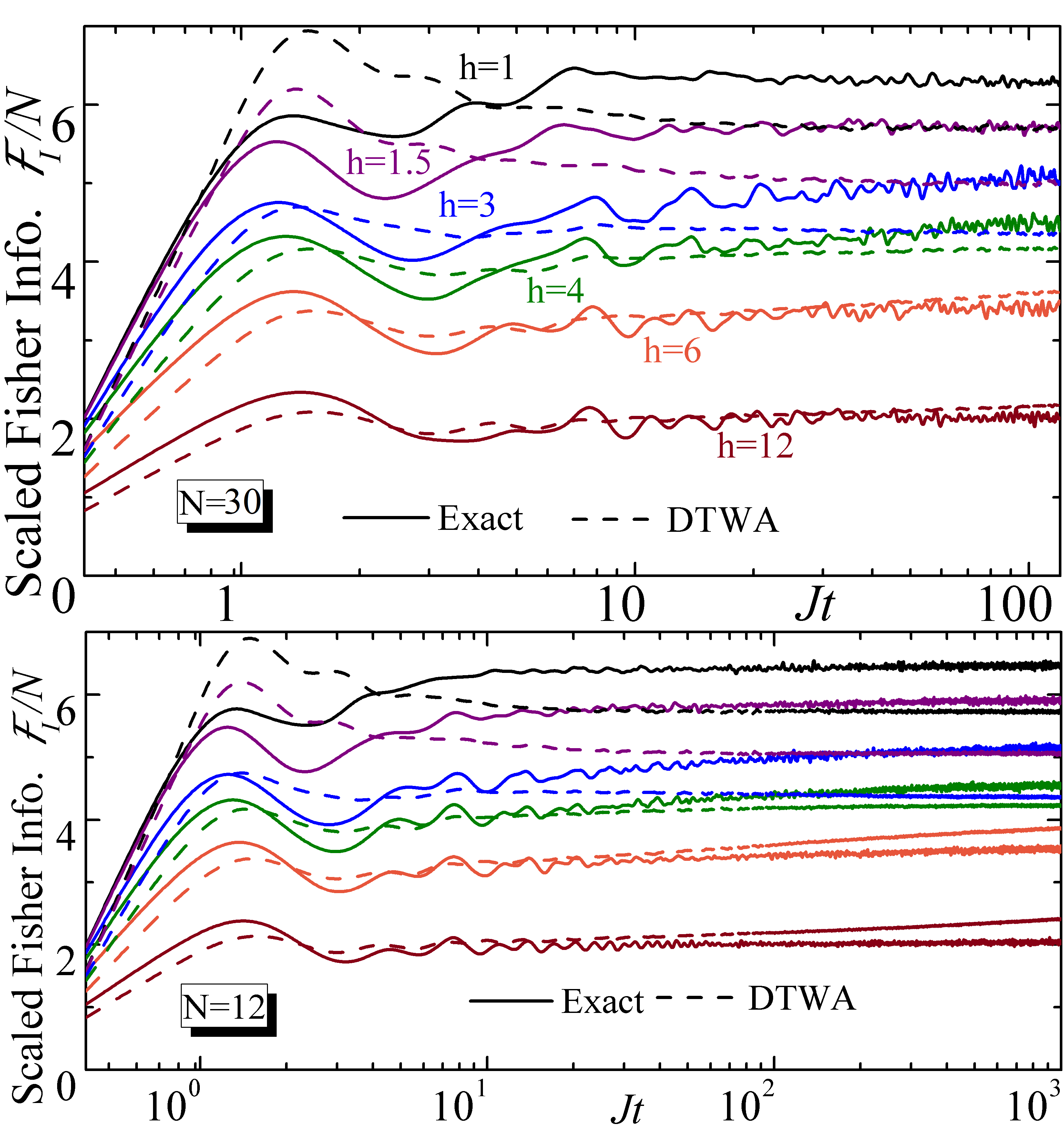}
  \caption{Evolution of scaled quantum Fisher information (QFI), which as measure of multipartite entanglement. Similar to Rényi entropy, and entanglement entropy, the localized phase is characterized by $\mathcal{F}_{I}\sim\log t$, a growth that breaks down in the thermal regime. Even for this measure, DTWA demonstrates its suitability for exploring the MBL transition as it also predicts both a logarithmic growth and a breakdown due to thermalization. System size $N=30$ (top) and $N=12$ (bottom). Exact numerical solutions obtained as in Fig. \ref{figimb}.} \label{figfish}
  \end{figure}

Contrary to imbalance and two-site Rényi Entropy, the QFI is a quantity that witnesses the build up of spin-spin correlations. Evidence of MBL has already been found using a QFI associated to the particle imbalance in a recent trapped ion experiment that simulated the 1D Ising model \cite{Monroe2016}. The fact that the long term behavior of entropy and QFI is logarithmic in time for the MBL phase is probably a coincidence restricted to short range models, and it may not be generally guaranteed. We observe the logarithmic growth of QFI in the DTWA results, showing a strong analogy of this semi-classical approach to the exact quantum system even to the level of spin fluctuations. However, a quantitative agreement between the exact and DTWA curves is now only possible in the large disorder limit, that is DTWA accurately captures the growth of QFI deep in the localized phase, but not deep in the thermal phase. Moreover the time scale over which DTWA accurately captures the growth of QFI grows longer as one goes deeper into the localized phase.%Only then spin-spin correlations can be well captured by an average over classical trajectories sampled accordingly to the initial probability distribution.\\

\begin{figure}[!t]
  \includegraphics[width=0.49\textwidth]{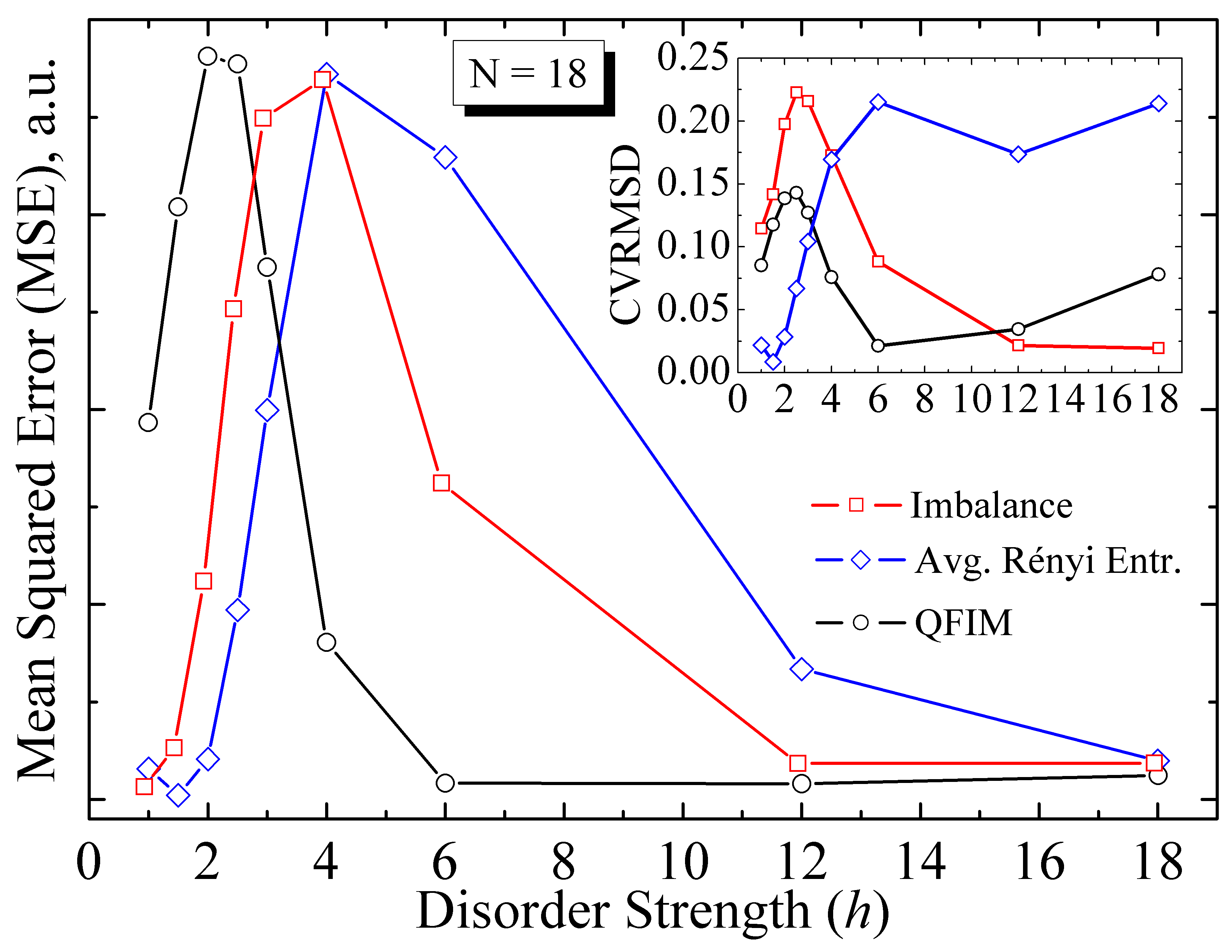}
  \caption{A summary of the agreement between the exact solution and DTWA results as a function of disorder strength in terms of Mean Squared Error (MSE, main figure) and Coefficient of Variation of the Root Mean Squared Deviation (CVRMSD, inset). More details in main text.
  %The overall tendency is clear, the error is small in the localized and thermal regime extremes and gets much bigger around the MBL transition.
  All MSE curves show a tendency to have bigger mismatches around the MBL transition and smaller errors deep in the localized and thermal phases.}
   \label{figdiff}
  \end{figure}

We present a summary of the suitability of the DTWA for reproducing several dynamical indicators of MBL in Fig. \ref{figdiff}. The curves correspond to a Mean Squared Error (MSE) of the form,
\begin{equation}
MSE = \frac{\sum_{i=1}^{M} (f_i-g_i)^2}{M}
\label{eqerror}
\end{equation}
where $M$ is the number of computed points of the quantities (which lie in the interval $Jt\in \pas{0,120}$), and $f_i$ ($g_i$) are the exact (DTWA) values. Although there are other ways to statistically analyze the discrepancy between exact and DTWA results, we have found that Eq. \ref{eqerror} is a very simple and common approach that express the visually qualitative closeness between the curves in the figures presented so far. The units of the MSE have been left unspecified as there are very different quantities displayed in the same figure, and what is relevant is the tendency of the agreement for each curve as the disorder strength varies. All MSE curves have maxima around the critical disorder strength and go to smaller values in both thermal and localized extremes. Other absolute error measures like Root Mean Squared Deviation or Mean Absolute Error behave in a similar fashion. We also included in the inset of Fig. \ref{figdiff} an example of a relative or normalized statistical measure of the error, namely the Coefficient of Variation of the Root-Mean-Squared-Deviation (CVRMSD), which is defined as $\varepsilon = \sqrt{MSE}/\bar{f}$, where $\bar{f}=\sum_{i=1}^M f_i$. The CVRMSD has the advantage of being dimensionless, so all curves can be put in the same scale. However, the CVRMSD has the disadvantage of being too sensitive to fluctuations in the exact solution with respect to the DTWA for quantities like the QFIM and Rényi Entropies in the localized phase, which remain very close to zero during the dynamics. We have already seen that DTWA is not well suited for reproducing those fluctuations, but only to show the average tendency of the relevant MBL observables. As it is the average tendency which is the most relevant aspect for MBL, we think that the MSE provides a better measure for the suitability of DTWA for exploring MBL dynamics.

\section{Final remarks}
\label{secconc}
%We have shown that the semi-classical picture provided by DTWA accurately models the quantum dynamics deep in both the localized and thermal phases, failing only close to the transition.
We have shown that the semi-classical picture provided by DTWA is able to reproduce the general features of the quantum dynamics in both the localized and thermal phases. The agreement persists up to time scales that grow longer as one goes deeper into either phase. Particular diagnostics we have considered include the memory of the initial conditions in imbalance, growth of R\'{e}nyi entropy for a two spin subsystem, and quantum Fisher information. While the discrepancy between exact methods and DTWA grows larger at long times, the DTWA works reasonably well over a range of intermediate timescales (which can be made longer by going deeper into the respective phases). It is worth noting that the DTWA can capture the main features of the dynamics over the range of timescales relevant for experiments e.g.~\cite{Schreiber2015}. %The correspondence is particularly relevant in the localized phase, where the semi-classical phase method also exhibits logarithmic growth of both entropy and QFI.
It is also worth noting that in this paper we have focused on the short-range 1D Heisenberg model, which is presumably the `worst case' scenario for a method of this type \cite{Schachenmayer2015,Schachenmayer2015a,Pucci2016,Pucci2017,orioli}. The DTWA is expected to become more accurate for problems with higher coordination number (either long range interacting problems or higher dimensional problems). These long range interacting and/or high dimensional problems are formidably difficult to tackle using exact techniques. However, %when the degrees of freedom become more connected, either by longer range interactions or spin lattices in high dimension. An important advantage is the ability of
when treated using DTWA, the computational cost scales only polynomially with system size. As such, we expect DTWA can become a powerful method for investigating localization and thermalization in these more complex and less explored settings, which are inaccessible to existing numerical techniques, but are of relevance for ongoing experiments. %Moreover the fact that experiments have limited coherence time makes the comparison even more valuable, since it is during such finite time windows when the approximation works better.

\begin{acknowledgements}

We thank Robert Lewis-Swan, Asier Piñeiro-Orioli, Shainen Davidson, and Anatoli Polkovnikov for insightful discussions. We acknowledge support from NSF-PHY-1521080, JILA-NSF-PFC-1125844, DARPA (W911NF-16-1-0576 through ARO), MURI-AFOSR and AFOSR. R.N. is supported by the Foundational Questions Institute (fqxi.org; grant no. FQXi-RFP-1617) through their fund at the Silicon Valley Community Foundation.

\end{acknowledgements}

\bibliography{References}

\end{document}